# IS THE ACCRETION FLOW IN NGC 4258 ADVECTION-DOMINATED?

J.-P. Lasota[1],

M. A. Abramowicz[2], X. Chen[2],

J. Krolik[3],

R. Narayan[4] and I. Yi[4]

## ABSTRACT

The mass of the central black hole in the active galaxy NGC 4258 (M106) has been measured to be $M = 3.6 \times 10^7 M_\odot$ (Miyoshi et al. 1995). The Eddington luminosity corresponding to this mass is $L_E = 4.5 \times 10^{45}$ erg s$^{-1}$. By contrast the X-ray luminosity of the nucleus of NGC 4258 between 2 − 10 keV is $(4 \pm 1) \times 10^{40}$ erg s$^{-1}$ while the optical/UV luminosity is less than $1.5 \times 10^{42}$ erg s$^{-1}$. The luminosity of NGC 4258 is therefore extremely sub-Eddington, $L \sim 10^{-5} L_E$ in X-rays and $L \sim 3 \times 10^{-4} L_E$ even if we take the maximum optical/UV luminosity. Assuming the usual accretion efficiency of 0.1 would imply accretion rates orders of magnitude lower than in Seyfert galaxies and quasars. We show that the properties of the AGN in NGC 4258 can be explained by an accretion flow in the form of a very hot, optically-thin plasma which advects most of the viscously generated thermal energy into the central black hole and radiates only a small fraction of the energy. In this case the accretion rate in Eddington units could be as high as $\sim 0.16\alpha$, where $\alpha$ is the standard viscosity parameter; and the size of the hot disk should be larger than $\sim 10$ times the Schwarzschild radius. We compare the predictions of this model with observations and discuss its application to other low luminosity AGN.

*Subject headings:* accretion, accretion disks — black hole physics — galaxies: individual (NGC 4258) — galaxies: nuclei

---

[1]UPR 176 du CNRS; DARC, Observatoire de Paris, Section de Meudon, 92195 Meudon, France

[2]Department of Astronomy and Astrophysics, Göteborg University and Chalmers University of Technology, 412 96 Göteborg, Sweden

[3]Department of Physics and Astronomy, the Johns Hopkins University, Baltimore, MD 21218, USA

[4]Harvard-Smithsonian Center for Astrophysics, 60 Garden Street, Cambridge, MA 02138, USA

## 1. INTRODUCTION

The spiral galaxy NGC 4258 (M106) has been classified (Stauffer 1982) as a LINER (i.e. a galaxy with a 'low ionization nuclear emission-line region'). Since LINERs are defined just by the properties of some line intensity ratios (e.g. Osterbrock 1989) (in the 'classical' definition by [OII]$\lambda$3727/[OIII]$\lambda$5007$\geq$ 1, and [OI]$\lambda$6300/[OIII]$\lambda$5007 $\geq$ 1/3) it is not surprising that they appear to form a heterogeneous class of objects (e.g. Filippenko 1993). Some LINERs however could contain 'dwarf' AGN and represent a 'missing link' between low luminosity AGN and nearby 'normal' galaxies. The best evidence of the presence of a 'dwarf' AGN is the detection of a pointlike non-thermal X-ray source at the nucleus, whose luminosity exceeds the Eddington limit for X-ray binaries (Makishima et al. 1994; Petre et al. 1993, Reichert, Mushotsky, & Filippenko 1994). The presence of radio/optical jets is also thought to be the signature of an 'accretion-ejection' structure in the nucleus.

'Dwarf' AGN are commonly supposed to accrete at rates lower than normal AGN, which would explain their low luminosities measured in Eddington units. The Eddington luminosity is one of the most fundamental parameters determining the properties of an accreting object. However, despite the importance of this quantity, accurate measurements have been very difficult to make in most AGN because, while the luminosity is usually comparatively easy to determine, there are very few clear indicators of the mass. Consequently, the first good measurement of $L/L_E$ in an AGN was published only this year (Miyoshi et al. 1995) for NGC 4258. In this case the motions of $H_2O$ maser spots permit a very accurate measurement of the central mass. Most of the uncertainty in the inferred $L/L_E \simeq 10^{-4.5\pm0.5}$ in this source is, in fact, due to the difficulty of guessing how much of the bolometric luminosity emerges in the infrared. The only other AGN with a relatively reliable mass determination (by the the Hubble Telescope) is M87 (Ford et al. 1994; Harms et al. 1994). Other good quality mass determinations are in 'normal' galaxies like M31 and NGC 3115 (Kormendy 1993).

The upper limit of $L/L_E < 10^{-4}$ implied by the observations in NGC 4258 is surprisingly small. If it is typical of all AGN, luminous quasars with bolometric luminosities $\sim 10^{47}$ erg s$^{-1}$ would have masses $\sim 10^{13} M_\odot$, or $\sim 100\times$ the mass of a typical galaxy. NGC 4258 must therefore have an unusually small value of $L/L_E$ relative to the rest of the AGN population. The fact that it is a LINER already marks it as different from more "classical" varieties of AGN, such as Seyfert galaxies and quasars. Many of the usual hallmarks of activity (e.g., strong non-stellar continuum in the optical/ultraviolet or broad emission lines) are absent in LINERs.

According to the standard, geometrically thin, accretion disk model, luminosities of the order of $(10^{-4} - 10^{-5})L_E$ should correspond to accretion rates $\dot{m} \equiv \dot{M}/\dot{M}_E \sim 10^{-3} - 10^{-4}$, assuming a 10% efficiency of accretion, where $\dot{M}_E = L_E/c^2 = 4\pi G M m_p/\sigma_T c$. The standard disk model cannot however explain the hard X-rays seen in AGN spectra, including LINERs and NGC 4258 in particular. The effective temperature at low accretion rates is (Frank, King, & Raine 1992)

$$T_{\text{eff}} \approx 3.5 \times 10^7 m^{-1/4} \dot{m}^{1/4} r^{-3/4} \text{K}, \tag{1}$$

and the optical depth is

$$\tau \approx 56\alpha^{-4/5}\dot{m}^{1/5}m^{1/5}, \tag{2}$$

where $m = M/M_\odot$, $r$ is the radius in units of the Schwarzschild radius $r_S = 2GM/c^2$, and $\alpha \leq 1$ is the Shakura-Sunyaev viscosity parameter. Here it is assumed that the radiation pressure is negligible and that the Rosseland mean opacity is approximated by the Kramers law. For parameters appropriate to NGC 4258 the standard inner disk would be cool ($T_{\rm eff} \sim 10^4$ K) and optically thick. It cannot emit hard (or soft) X-rays. For $m = 10^7$ and $\dot{m} = 10^{-4}$ say, properties of AGN accretion disks are close to those of accretion disks in cataclysmic variables (Eq. [1] and [2]). In cataclysmic variables however, if (weak) X-rays are observed at all, they originate in a boundary layer (Córdova 1995) so that it is not clear how, in the AGN case, X-rays could be produced in the framework of the standard thin disk model. Neufeld & Maloney (1995) did, however, explain some of the properties of the outer, molecular disk in NGC 4258 in a model with $\dot{M} = 7 \times 10^{-5}\alpha M_\odot$ yr$^{-1}$ and a warped disk.

In order to explain the X-rays emitted by Seyfert galaxies and quasars it is now common to assume the existence of a higher temperature region in the vicinity of the accretion disk. For example, in the disk-corona models, the hot layer upscatters soft photons from the cold layer to produce the X-ray power-law, and the UV black body radiation is formed in the cold layer (Haardt & Maraschi 1993). Such models should produce a characteristic spectral 'signature' in the form of a bump in the spectrum between 10 and $\simeq 30$ keV due to Compton reflection off the optically thick accretion disk, and an Fe K$\alpha$ line with equivalent width $\sim 100$ eV (Lightman & White 1988; George & Fabian 1991).

Here, we consider the possibility that hard X-ray emission at very sub-Eddington luminosities may arise via a model which is very different from the 'standard' one. In section 2 we show that the X-ray spectrum of NGC 4258 can be reproduced by an advection-dominated flow. Section 3 discusses the relevance of advection-dominated flows to other low luminosity AGN. In section 4 we test the model further by using the predicted spectrum to calculate the narrow optical emission line ratios. Finally, we conclude in section 5 with a brief discussion.

## 2. ADVECTION-DOMINATED FLOWS IN NGC 4258

In applying 'non-standard' accretion flow models to low-luminosity systems, i.e. models in which no standard (cold) disk is present close to the compact object, there are essentially two possibilities to explain the low luminosity.

One is that the accretion rate is indeed very small and the black hole is 'starved' (Rees et al. 1982). This model has been invoked to explain some properties of LINERs but in its framework X-ray spectra do not find a natural explanation (Petre et al. 1993).

The second possibility is that the black hole accretes at 'reasonable' rates but very little of the gravitational energy released by accretion is radiated (Narayan & Yi 1994; Narayan & Yi

1995ab; Abramowicz et al. 1995; Chen et al. 1995). When the latter possibility applies, the disk is said to be "advection-dominated".

The "advection-dominated" disk model has been successfully applied to low-luminosity X-ray sources such as Sgr A* (Narayan, Yi, & Mahadavan 1995) and quiescent 'soft X-ray transients' (Narayan, McClintock, & Yi 1996).

The photon spectral index above 3 keV in NGC 4258 is $\Gamma = 1.78 \pm 0.29$, which is typical of AGN (Nandra & Pounds 1994). Unfortunately the absorbing column is large, $1.5 \pm 0.2 \times 10^{23} \mathrm{cm}^{-2}$, which means that there is no way to directly search for any optical/ultraviolet continuum from the nucleus of NGC 4258. If there were a strong continuum in this wavelength range, we might be able to infer its presence by detecting its luminosity reradiated by the obscuring dust in the mid-infrared. However, as of yet there is no mid-infrared imaging of this object with the necessary angular resolution. The only evidence available at this time is a large aperture upper limit on the mid-infrared flux, $1.5 \times 10^{42}$ erg s$^{-1}$ (Rieke & Lebofsky 1978). Much of this infrared flux is likely to be reprocessed starlight, so this upper limit is probably considerably greater than the true nuclear infrared flux.

We have used a two-temperature model of advection-dominated accretion onto a black hole (Narayan & Yi 1995b, Narayan et al. 1995) to model the observed properties of the continuum emitted by the nucleus of NGC 4258. The accretion flow in our model consists of an inner advection-dominated flow and an outer 'standard' geometrically thin disk. In the advection-dominated region, the ion temperature is close to the virial temperature, while the electron temperature is determined by a balance between heating due to collisions with the ions and inverse Compton cooling. Typically, we find $T_e > 10^9$K. We also assume that there is a magnetic field whose energy density is in equipartition with the ion pressure. The synchrotron photons radiated by the hot electrons provide seeds for inverse Compton scattering.

The properties of the output spectrum depend on the black hole mass $m$ (fixed by observations), the viscosity parameter $\alpha$, the ratio of gas to total pressure $\beta$, the accretion rate $\dot{m}$, and the inner and outer radii of the advection-dominated zone and the thin disk. The results depend significantly on only two parameters, $\dot{m}/\alpha$ and the inner radius of the cold thin disk $r_{\mathrm{in}}$. The former quantity fixes the absolute luminosity and weakly influences the X-ray slope, in the sense that larger values of $\dot{m}/\alpha$ gradually lead to harder spectra (Narayan 1996). The latter determines the contribution of standard disk emission (in the optical/ultraviolet) to the total spectrum. When $r_{\mathrm{in}}$ is as small as $\simeq 10$, photons from the standard disk also contribute significantly to the seed population for inverse Compton scattering in the hot advection-dominated part of the disk, and the X-ray spectrum becomes significantly softer; when $r_{\mathrm{in}}$ is larger than this, the seed photons are dominated by locally-produced synchrotron photons and the X-ray spectrum is independent of $r_{\mathrm{in}}$.

Since the optical upper limit is not very constraining, we fitted the spectrum of NGC 4258 with four models differing by the contribution of the standard disk to the total emission. The results are shown in Fig. 1.

In all four models, $m = 3.6 \times 10^7$, $\alpha = 0.1$, $\beta = 0.95$, and $\dot{m} = 0.016$. As already mentioned, the luminosity depends only on $\dot{m}/\alpha$, and this quantity has been adjusted so as to fit the observed X-ray luminosity of NGC 4258. The inner edge of the advection-dominated flow is fixed at $r = 3$, and the outer edge of the outer thin disk at $r_{\text{out}} = 10^6$. In the first model there is no outer cold disk, while in the others $r_{\text{in}} = 1000$, $r_{\text{in}} = 100$ and $r_{\text{in}} = 10$ respectively. Decreasing $r_{in}$ gives increasing optical/UV emission. The model with the standard disk extending down to $r_{\text{in}} = 10$ does not fit the X-ray spectrum since the advection-dominated flow is too efficiently cooled by the soft photons from the standard disk. The other three models all have slopes in the X-ray band consistent with the one observed (see Fig. 1).

Thus, with $\dot{m}/\alpha$ set by the X-ray luminosity, this model predicts an X-ray spectral slope which lies within the error bar of the observed value. While the advection-dominated model is thus *consistent* with what we know about NGC 4258, it is not however a *unique* explanation. First of all, the uncertainty in the measured slope is substantial. Second, any thermal Comptonization model in which the seed photon luminosity is $\sim 0.1\times$ the cooling rate of the hot plasma yields the correct spectral shape in the X-ray band (Haardt, Maraschi, & Ghisellini 1994; Pietrini & Krolik 1995; Stern et al. 1995), as do also models in which the seed luminosity to cooling ratio is $\simeq 0.5$ and the Compton depth is slightly less than 0.1 (Haardt & Maraschi 1993; Zdziarski et al. 1994). What distinguishes the advection-dominated model is that it predicts this spectrum with a minimum of parameter-tuning. Only two assumptions are required: $r_{\text{in}}$ must be large enough that the normal disk does not provide significant seed photons to the hot disk, and the magnetic field must be near equipartition with the gas pressure, as is suggested by the simulations of Hawley, Gammie, & Balbus (1995).

Another constraint on these models can be found from polarization studies revealing the nuclear optical continuum via external reflection. For example, on the basis of this kind of data, Wilkes et al. (1995) find that the spectral shape of the nuclear optical continuum is roughly $F_\nu \propto \nu^{-1.1}$, and estimate that its luminosity is $\sim 10^{40} - 10^{41}$ erg s$^{-1}$. Such a spectral shape and luminosity would be in very good agreement with our models when $r_{in}$ is greater than a few hundred.

Superimposed on the X-ray continuum, there is also an Fe K$\alpha$ emission line of EW$\simeq$ 250 eV (Makishima et al. 1994). In order to produce a fluorescent line with an equivalent width this large, somewhat more than half of the solid angle around the continuum source must be covered with a gas shell whose surface density is $\gtrsim 10^{24}$ cm$^{-2}$. While this line might be taken as evidence for a nearby cool accretion disk, it could also be explained by fluorescence in an obscuring torus, provided the mean column density is roughly an order of magnitude greater than the column density on our line of sight, $1.5 \times 10^{23}$ cm$^{-2}$.

## 3. NGC 4258 AS A REPRESENTATIVE LINER?

Where does our model place NGC 4258 within the AGN 'world'? Is it a 'missing link' between low luminosity AGN and 'normal' galaxies? In our model, the accretion rate in Eddington units is $\dot{m} = 0.16\alpha$, which gives $\dot{m} = 0.016$ for a standard $\alpha$ of 0.1. As mentioned in §1, it is not easy to tell what the Eddington luminosity is in other AGN. However, the very crude indications we have suggest that $L/L_E$ may be systematically smaller in LINERs than in other AGN classes. As we remarked above, $L/L_E$ in quasars cannot be as small as in NGC 4258. If, as is widely believed, the reflection regions in Seyfert galaxies are winds, they must have $L/L_E$ at least $\sim 0.1$ (Krolik & Begelman 1986, Balsara & Krolik 1993). On the other hand, when attempts are made to estimate $L/L_E$ in LINERs, very small numbers are found: $\simeq 10^{-4.5}$ in the case of NGC 4258, and on the basis of much less direct evidence, $\sim 10^{-3.5\pm1}$ in M81 (Ho, Filippenko, & Sargent 1995). It is therefore tempting to speculate that the parameter which distinguishes LINERs from other AGN is their small $L/L_E$, as in NGC 4258.

Ten other LINERs have been observed in X-rays (Petre et al. 1993; Reichert et al. 1994; Makishima 1994; Ishisaki 1994; Terashima et al. 1994; Koratkar et al. 1995): M 81, NGC 7213, NGC 3998, NGC 4594, NGC 4579, NGC 2639, NGC 3079, NGC 3642, NGC 4278, and Pictor A. Their spectra are quite diverse, but their (photon) slopes in the 2 – 10 keV band are generally $\simeq 1.7$ or steeper. In the best observed case, the nucleus of M 81, Petre et al. (1993) found a slope of 2.2 above 1 keV, but Ishisaki (1994) argued that the earlier observations were contaminated by extra-nuclear sources, and $\Gamma$ of the nuclear source is really $\simeq 1.7$. At the same time, the X-ray flux is relatively strong compared to the ultraviolet flux. The mean value of the slope of a power-law interpolated between the ultraviolet and the X-rays is $\simeq 1.1$ in LINERs, while it is $\simeq 1.4$ in Seyfert galaxies (Petre et al. 1993).

The relative strength of the X-rays compared to the ultraviolet flux for these other LINERs is very much in keeping with the predictions of the advection-dominated model. Some of the other LINERs also have X-ray slopes similar to the one observed in NGC 4258, and therefore consistent with our model. However, to accommodate those with steeper slopes while retaining relatively weak ultraviolet emission requires reducing $\dot{m}/\alpha$ roughly an order of magnitude below the value in NGC 4258 (Narayan 1996).

It is also interesting to compare the advection-dominated model of NGC 4258 with that of the radio source in the center of our Galaxy Sgr A* (Narayan et al. 1995). The X-ray luminosity of this source $L_X \lesssim 10^{-7} L_E$ is more than two orders of magnitude lower than in NGC 4258. The accretion rate according to the advection-dominated disk model is however only $\sim 20$ times lower: $\dot{m} = 7.84 \times 10^{-4}$ (with $\alpha = 0.1$). This is typical of advection-dominated flows in which the luminosity roughly scales with the square of the accretion rate (Narayan et al. 1996). One can speculate that if LINERS are the 'missing link' between AGN and normal galaxies the accretion rates in LINERs should typically be $\dot{m} \sim 10^{-2}\alpha$.

## 4. EMISSION LINES

Finally, some information about the nature of the nucleus of NGC 4258 may be provided by the narrow emission lines present in the optical spectrum. Considering LINERs as a class, it was suggested some years ago that the line ratios could be explained by a steep power-law UV/X-ray continuum photoionizing a gas shell around the nucleus under conditions of low ionization parameter (Halpern & Steiner 1983; Ferland & Netzer 1983). A 7 - 8×10$^4$ K black-body plus a flat soft X-ray spectrum was also suggested (Péquignot 1984). Other work has claimed that the particular line ratios seen in NGC 4258 in the gas within a few arcsec (30 − 150 pc) of the nucleus are better explained by photoionization due to an ultraviolet-dominated continuum with a spectral index of 1.2 and a total UV luminosity of $1.3 \times 10^{42}$ erg s$^{-1}$ (Stüwe, Schulz, & Hühnermann 1992). If there were no obscuration, an ultraviolet continuum of this strength would exceed the IUE flux by about an order of magnitude (Ellis et al. 1982). It is also very nearly as large as the upper limit on the bolometric luminosity placed by the observed 10$\mu$ luminosity within 90pc of the galaxy's center, $1.6 \times 10^{42}$ erg s$^{-1}$ (Rieke & Lebofsky 1978).

One should remark however that interpretation of these observations is made delicate by the existence of line ratio gradients which may not be completely resolved, even with the 0.5" resolution obtained by Stüwe et al. (1992). *HST* observations could help clarify the situation.

To test whether the ionizing continuum shape can be inferred from the relative strengths of the narrow emission lines, we have run a number of photoionization models using the code XSTAR (Kallman & Krolik 1993), assuming the continuum has the spectrum shown in Fig. 1 (from the models with no outer cool disk). In contrast to most photoionization-modelling, for which the ionizing flux, the continuum spectral shape, the gas pressure, and the thickness of the clouds are all free parameters, in this case only the last two are free: the advection-dominated model determines both the ionizing luminosity and its spectral shape, while the observations of Stüwe et al. (1992) determine the distance of the clouds from the nucleus.

We studied a spherical shell of inner radius 18 pc. When the gas pressure is $3 \times 10^{-10}$ dyne cm$^{-2}$ (i.e. the H nucleus density varies between 60 and 200 cm$^{-3}$), and the cloud thickness is in excess of a few $\times 10^{21}$ cm (i.e. the shell is at least half a dozen pc thick), we find a good match between some of the predicted line ratios and those observed. The observed line luminosity requires this shell to cover most of the solid angle around the nucleus.

With these parameters, the predicted H$\alpha$ luminosity is $\simeq 1.1\times$ the observed H$\alpha$ luminosity within 4″ of the nucleus (Stauffer 1982). The flux ratio [OI]6302/H$\alpha \simeq 1.5$ in the model, while the actual number is $0.5 - 1.7$, depending on location (Stüwe et al. 1992). The [NII]6585 flux predicted by the model is the same as H$\alpha$ to within 10%, just as in the data, while [SII]6716,6731/H$\alpha$ is 0.4 in the model, versus the observed value of 0.6. Note that we have here compared to the line ratio data of Stüwe et al. (1992); the flux ratios found by Stauffer (1982) sometimes differ at the factor of two level. A similar quality fit can be found using a conventional spectrum (Stüwe et al. 1992).

There are, however, discrepancies between the observations and the line strengths predicted both by the advection-dominated spectrum and by a conventional power-law spectrum. In particular, both predict about 10 − 15 times more [OIII]5007 than is reported (Stauffer 1982). It is possible that the line ratios in NGC 4258 cannot be explained by a single zone model. Clearly a higher quality, homogeneous, set of data is required to answer this question.

## 5. DISCUSSION AND CONCLUSIONS

NGC 4258 also shows other signs of activity that could be related to the presence of an advection-dominated flow. Helically twisted, nuclear jets are observed in radio, visual and X-ray radiation (Cecil et al. 1995). Advection-dominated disks could produce these jets by generating magnetically-driven bipolar outflows near the axis of the quasi-spherical accretion flow (Narayan & Yi 1995a; Rees et al. 1982).

We have shown that the X-ray spectrum of NGC 4258 can be explained by the presence of a hot advection-dominated flow. The observed luminosity helps us fix $\dot{m}/\alpha = 0.16$, and with this choice our model correctly predicts the slope of the X-ray spectrum so long as $r_{\rm in}$ is significantly greater than 10. The lack of data in other wavelengths, in particular in the mid-infrared, does not allow for more detailed tests of the model at this time. The example of NGC 4258 and other LINERs shows how important the determination of the black hole mass is for comparing models to observations.

We thank Suzy Collin and Daniel Péquignot for very helpful remarks. XC thanks the Observatoire de Paris, Section de Meudon for hospitality during July 1995 when this work was in the final stage. RN was supported in part by NSF grant AST 9423209. JK and RN thank the Institute for Theoretical Physics, Univ. of California, Santa Barbara, for hospitality and support through NSF Grant PHY9407194.

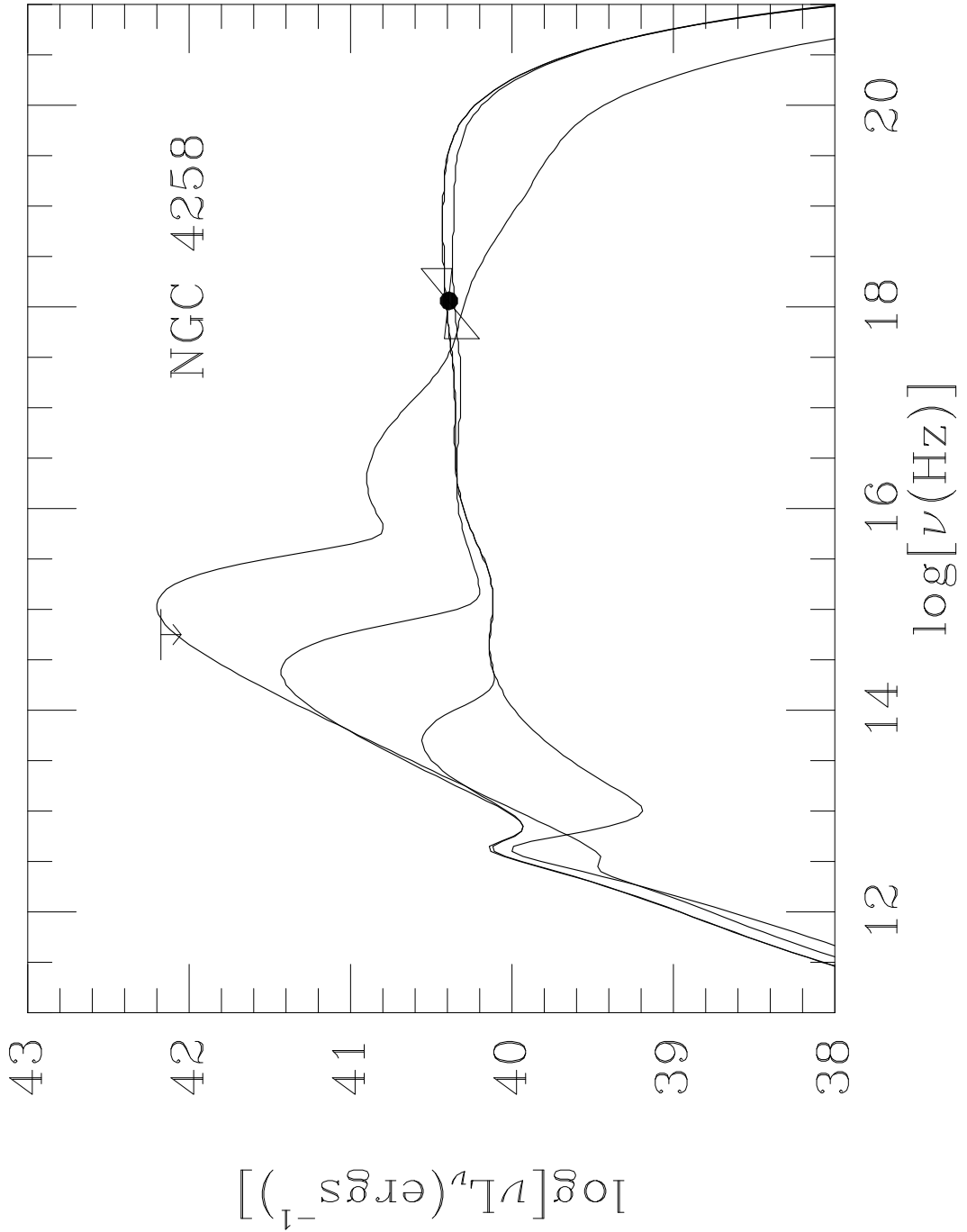

Fig. 1.— Spectra of NGC 4258 calculated with an advection-dominated flow extending from $r = 3$ to $r = r_{in}$, and a standard 'cold' disk extending from $r = r_{in}$ to $r_{out} = 10^6$. The parameters are $m = 3.6 \times 10^7$, $\alpha = 0.1$, $\beta = 0.95$, and $\dot{m} = 0.016$. Four models are shown: (i) No outer cold disk ($r_{in} = r_{out}$), (ii) $r_{in} = 1000$, (iii) $r_{in} = 100$, (iv) $r_{in} = 10$. As $r_{in}$ decreases the optical/UV emission increases. The upper limit of $1.5 \times 10^{42}$ erg s$^{-1}$ in the optical is indicated. The bow-tie represents the measured flux and the 1 $\sigma$ range of slopes (1.49 to 2.07) of the X-ray spectrum measured by Makishima et al.(1994).